# Computational Assessment of Hemodynamics in Asymmetric-type Lesion of Idealized Coronary Stenoses

Ayodele Oyejide [a*], Oluwatosin Abodunrin [a, b], Ebenezer Ige [c] Adetokunbo Awonusi [d]

**Ayodele Oyejide**

[a] Department of Biomedical Engineering, College of Engineering, Afe Babalola University, Ado-Ekiti, Nigeria, 360231 aoyejide@abuad.edu.ng

**Oluwatosin Abodunrin**

[b] Euromed Research Center, Euromed Engineering Faculty, Euromed University of Fes abodunrinod@abuad.edu.ng

**Ebenezer Ige**

[c] Department of Mechanical Engineering, Rochester Institute of Technology, NY-14623 USA eoieme@rit.edu

**Adetokunbo Awonusi**

[d] Department of Mechanical and Aerospace Engineering, Oklahoma State University Stillwater, OK-74078

adetokunbo.awonusi@okstate.edu

**\*Corresponding authors emails**: aoyejide@abuad.edu.ng

## Abstract

Coronary artery stenosis, characterized by the narrowing of the lumen, significantly affects blood flow and contributes to the progression of cardiovascular diseases. This study investigates the hemodynamics of coronary artery models with varying stenosis configurations, all maintaining an 80% lumen reduction, to determine how differences in morphology influence flow behavior and mechanical stresses. We employed computational fluid dynamics to analyze five idealized geometries with (10% & 70%), (20% & 60%), (30% & 50%), (40% & 40%), and (0% & 80%) stenosis configurations. Through physiological pulsatile flow conditions, we evaluated key hemodynamic pattern including velocity profiles, wall shear stress, and pressure distribution. Our results reveal that despite the same degree of lumen reduction, each stenosis configuration produced distinct flow patterns and hemodynamic profiles. Asymmetric configurations, such as 10% & 70% and 20% & 60%, exhibited pronounced flow disruptions and higher wall shear stress at the stenosis throats, while symmetric configurations, such as 40% & 40%, demonstrated more uniform flow and reduced vortex. Our findings challenge the practice of generalizing results across stenosis configurations without accounting for morphological variations, which is prevalent in many CFD studies using idealized models. This study emphasize the importance of considering stenosis-specific morphology in CFD analyses and clinical interpretations to enhance the accuracy of diagnostic tools, improve personalized treatment planning, and guide the design of medical devices such as stents.

## 1. Introduction

Cardiovascular issues, particularly the configuration and progression of atherosclerosis, represent a significant global health concern. This condition results in the narrowing or blockage of coronary arteries, disrupting healthy blood flow in the cardiovascular system. Numerous scientific studies, spanning molecular biology and engineering, have sought to address health challenges associated with atherosclerosis, including ischemic heart disease [1], [2], [3]. At the engineering level, biomedical researchers leverage numerical tools such as computational fluid dynamics (CFD) and fluid-structure interaction analyses [4], along with mathematical modeling [5], [6] to investigate critical parameters like blood flow velocity, pressure distribution, and wall shear stress. These studies often rely on models of the coronary arteries or the cardiovascular system developed from high resolution imaging techniques, including CT and MRI scanners [7]. This approach has enabled researchers to work with realistic models of human coronary arteries and develop idealized models to mimic the structure and behavior of real arteries or organs. However, while accurate physics and boundary conditions are essential for reliable simulations, the precision of computational results critically depends on the accuracy of the geometric models used [8], [9], [10].

Researchers have used idealized models of the coronary artery in several studies, often representing them as rigid pipes with truncated ends, due to the complexity and ethical issues involved in generating realistic models [11], [12], [9]. They place constrictions on specific areas to depict stenosis. One key advantage of using idealized geometries in biomedical research is the freedom it gives researchers to design stenosis constrictions (or plaque curvatures) of their choice. This flexibility makes it possible to hypothesize about possible physiological scenarios in the human body.

For instance, Carvalho et al. [13] compared hemodynamic results for CFD and FSI in uniform 50% stenosed idealized models of the coronary artery. They found only minor differences in the results, noting that the computational behavior closely aligned with the findings of chaichana et al. [14], who used CFD to study flow behavior in 60% coronary reductions across two left coronary artery models with plaques of equal mass on both walls. Their simulation showed that velocity and WSS peaked at the stenosis throat. kamangar et al. [15] analyzed hemodynamics in coronary arteries with 70%, 80%, and 90% lesions under steady and transient conditions using three different geometric shapes: elliptical, triangular, and trapezoidal. They demonstrated that the influence of geometry shape directly correlated with the intensity of stenosis. Dwidmuthe et al. [16] investigated how geometry and stenosis severity affected coronary arteries by studying lesions ranging from minor to severe (10%, 25%, 50%, 75%, and 90%) in an idealized model using large eddy simulation. Their results revealed robust trends consistent with other studies, showing that severe lesions (70–90%) produced unhealthy hemodynamic behavior. Freidoonimehr et al. [17] developed idealized geometries with more complex stenosis configurations, such as round, oval, elongated, half-moon, bean-shaped, and crescent-shaped geometries, with and without eccentricity, maintaining a constant 73% lesion in all cases. By comparison, their numerical results indicated that stenosis shapes did not significantly alter the hemodynamic outcomes, contrasting with the findings of Sarfaraz et al. [15]. Kabir et al. [18] conducted numerical simulations on symmetric 60% and 75% idealized stenosed arteries and observed notable increases in velocity, pressure drop, and wall shear stress as stenosis severity increased. Several other studies have specifically evaluated how lesion or stenosis curvature influences hemodynamics in idealized coronary arteries using CFD analyses [19, 20, 21, 22, 23].

Although various studies in the literature have examined similar degrees and configurations of idealized stenosis or lesions—particularly using concentric and eccentric shapes—simulation results for basic computational variables such as velocity, pressure, and wall shear stress often vary. Researchers frequently overlook the impact of different morphology on the same degree of lumen reduction or stenosis, leading to generalized conclusions regardless of stenosis shapes. It is crucial to exercise caution when defining flow boundary conditions for idealized geometries, as these significantly influence simulation outcomes and, more importantly, the interpretation of results for clinical and research applications [19, 10].

In this study, we critically analyzes hemodynamic behavior and its implications in idealized coronary artery geometries with an equal degree of lumen reduction but varying concentric ratios, such as used in the literature to assess whether generalizing outcomes is valid. Since stenosis conditions become critical at 80% lumen reduction [4], all geometries in this study were designed with an 80% area reduction. Transient flow simulations were performed on all models using computational fluid dynamics. Computational variables including blood flow velocity, wall pressure distribution, wall shear stress (WSS), and their respective volumes as blood flowed through the fluid domain—were obtained, analyzed, and compared to derive robust conclusions. The results revealed that each stenosis configuration produced distinct flow behaviors, pressure distributions, and WSS patterns.

The remainder of this paper is structured as follows: Sec. 2 presents the materials and methods, Section 3 details the results, discussions are provided in Section 4, and Section 5 concludes with suggestions for future work.

## 2. Materials and Methods

### 2.1. Geometry modeling

We created the idealized geometries using the SpaceClaim platform within the ANSYS Fluent 2022 software. Five distinct long models, each 49 mm in length and 4 mm in diameter, were designed with 80% eccentric stenoses positioned 26.8 mm upstream to mimic truncated segments of the coronary artery, as depicted in Figure 1. The upstream placement of constrictions allowed the blood flow to fully develop before encountering the stenosis. As illustrated in Figures 1 and 2, we imposed concentric stenoses on opposite ends of only four geometries, with lesion ratios of 10% & 70%, 20% & 60%, 30% & 50%,and 40% & 40% representing the less severe and most severe faces, respectively. The fifth model featured a one-sided stenosis with 80% reduction. For clarity, we denoted each stenosis larger radius R and the smaller one with r (Table 1). Table 1 provides the full lengths of the constrictions for R and r, respectively. In the fifth geometry, we applied a radius depth of 3.2 mm, which was varied for all larger constrictions throughout the study.

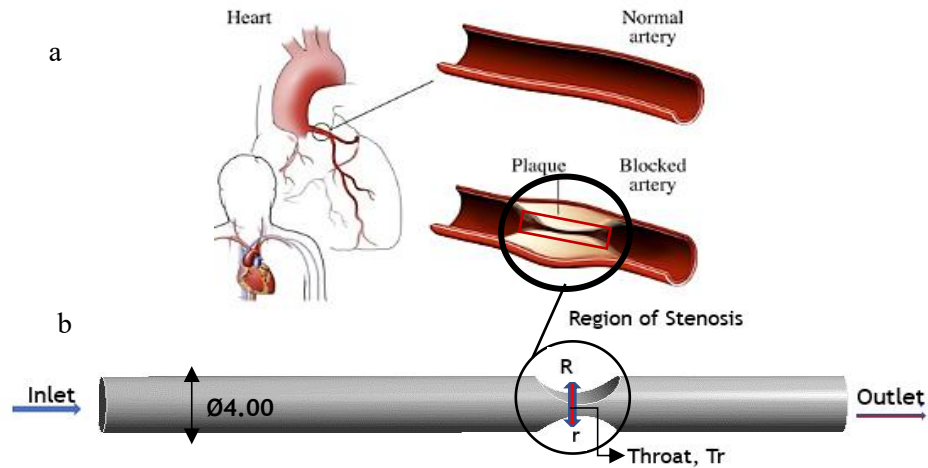

**Figure 1**. (a) Morphological depiction of a healthy and stenosed human coronary artery (b) An idealized model of a stenosed geometry, highlighting the lesion region.

**Table 1.** Basic parameters for the five (5) geometries

| Constriction degrees (%) | Smaller radius $r$ (mm) | Larger radius $R$ (mm) | Throat $T_r$ (mm) | Length $L$ (mm) | Diameter $D$ (mm) |
|---|---|---|---|---|---|
| 10 &70 | 3.1 | 6.3 | 0.8 | 49.0 | 4.0 |
| 20 & 60 | 4.2 | 6.1 | 0.8 | 49.0 | 4.0 |
| 30 & 50 | 5.1 | 5.9 | 0.8 | 49.0 | 4.0 |
| 40 & 40 | 5.5 | 5.5 | 0.8 | 49.0 | 4.0 |
| 80 | - | 6.4 | 0.8 | 49.0 | 4.0 |

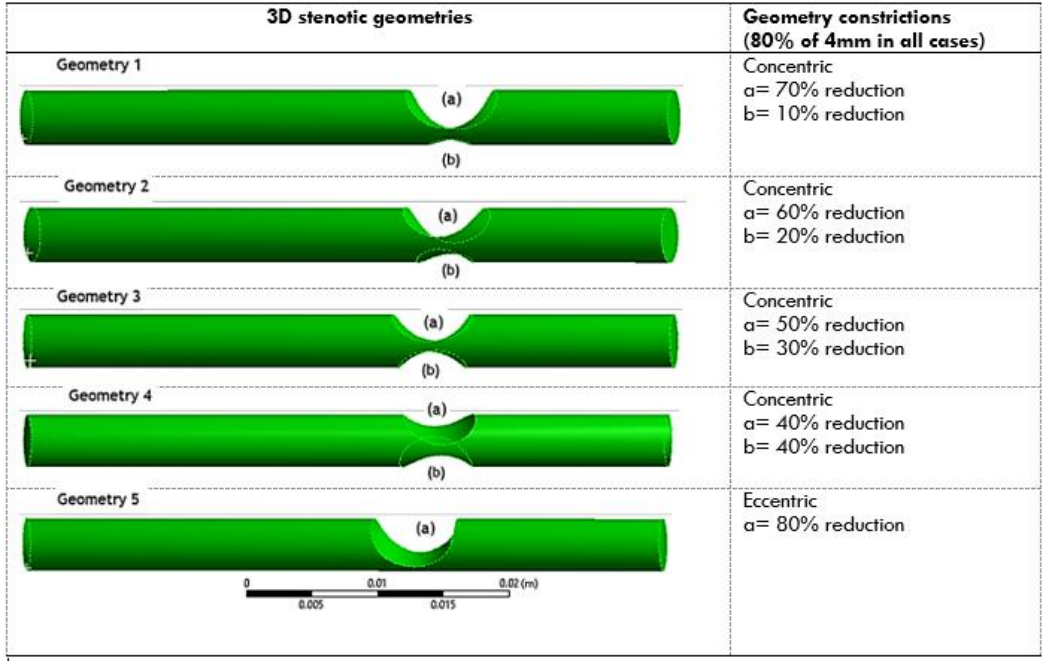

**Figure. 2.** Idealized 3D geometry variants used for the simulations. All variations result to 80% reduction in blood passage (3.2 mm of 4mm).

## 2.2. Generation of mesh

The developed coronary geometry models were prepared for meshing using the same ANSYS Fluent simulation setup. The tetrahedral mesh values, determined after conducting mesh studies, are provided in Table 2. A fine mesh was generated with element sizes of 0.0003 m applied to all models (see Figure 3), specifically to capture the velocity, pressure distribution, and wall shear stress. In total, 177,114 nodes and 932,450 elements were generated across all five models. To enhance the accuracy of the simulation at the constriction sites, inflation properties were applied, with a smooth transition ratio of 0.3 and a growth ratio of 1.2 in these regions.

**Table 2.** Mesh statistics for the five different geometries

| Constrictions (%) | Nodes | Elements | Element size (mm) |
|:---:|:---:|:---:|:---:|
| 10 & 70 | 35124 | 182350 | 0.0003 |
| 20 & 60 | 35620 | 187674 | 0.0003 |
| 30 & 50 | 35738 | 188318 | 0.0003 |
| 40 & 40 | 35968 | 189467 | 0.0003 |
| 0 & 80 | 34664 | 184641 | 0.0003 |
| **Total** | **177114** | **932450** | **0.0015** |

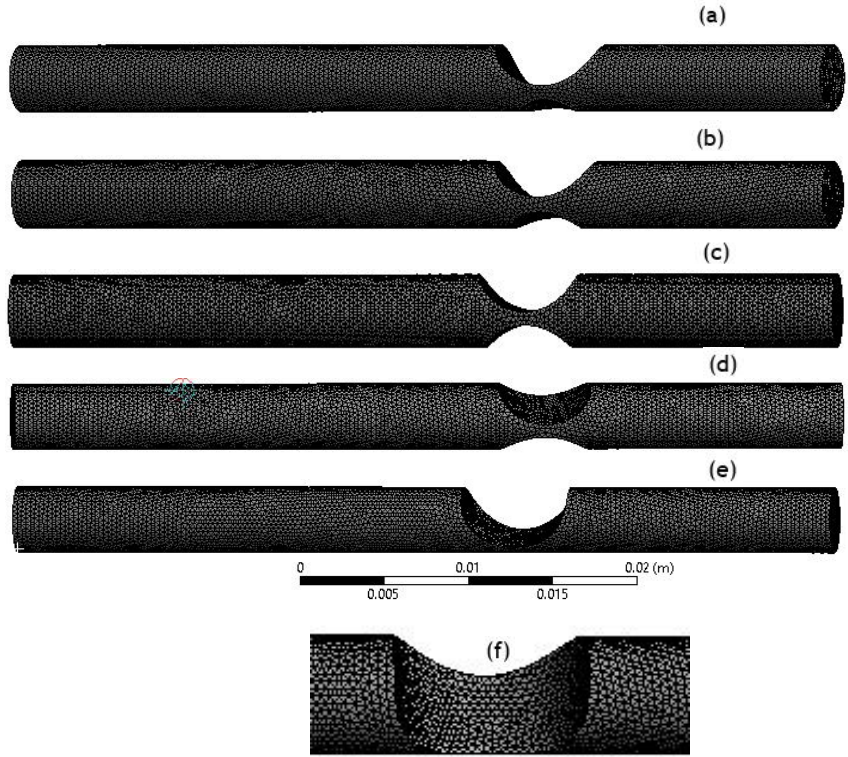

**Figure 3**. Meshed output for all five geometries in (a) 10% & 70% (b) 20% & 60% (c) 30% & 50% (d) 40% & 40% and (e) 80%, respectively. (f) Magnified view of the mesh fineness on the constrictions.

*2.3. Boundary conditions*

      To set the boundary conditions, we specified the inlet at the circular end of the longest distance on the geometries (upstream) and selected the truncated end of the shortest distance as the outlet (downstream). To enable independent studies of the fluid dynamics in both the non-stenosed and stenosed sections, we applied a no-slip condition on their respective walls, defined as *Wall₁* for the entire external body (excluding the inlet, outlet, and constriction surfaces) and *Wall₂* for the constricted surfaces. Additionally, the inlet velocity was defined as a periodic function using a user-defined function (UDF), ranging from 0.1 m/s during diastole to 0.5 m/s during contraction, as shown in the waveform in Figure 4.

*2.4.    Blood modelling*

Realistically, blood is a non-Newtonian incompressible fluid with pulsatile velocity. The non-Newtonian and incompressible components are generally accounted for by the continuity and Navier-Stokes equations represented by Equation. 1 and Equation. 2:

$$\nabla . v = 0 \qquad\qquad\qquad (1)$$

$$\rho \left( \frac{\partial v}{\partial t} + v . \nabla v \right) = - \nabla p + \mu \nabla^2 v + f \qquad\qquad (2)$$

In this study, we simulate pulsatile flow within the models by incorporating physiologically realistic, time-dependent boundary conditions. While the vessel walls were assumed to be rigid, the simulations captured key aspects of unsteady hemodynamic behavior associated with pulsatile blood flow (Section 3). We accounted for the realistic physiological blood shear-thinning behavior using the Carreau viscosity model, which is well-suited for modeling blood's non-Newtonian behavior under physiological conditions [22], [24]. The mathematical representation for this model is given in Equation 3 as:

$$\mu = \mu_\infty + (\mu_0 - \mu_\infty)[1 + (\lambda\gamma)^2]^{(n-1)/2} \tag{3}$$

Where, infinity shear viscosity, $\mu_\infty = 0.00345$ Kg/m$^{-2}$, zero shear viscosity, $\mu_0 = 0.056$ Kg/m$^{-2}$, time constant, $\lambda = 3.313$s, power law index, $n = 0.3568$, and $\gamma$ is the strain rate. These variables were inputted during the computational modelling set-up, which automatically assigns blood's non-Newtonian flow properties to the fluid under study.

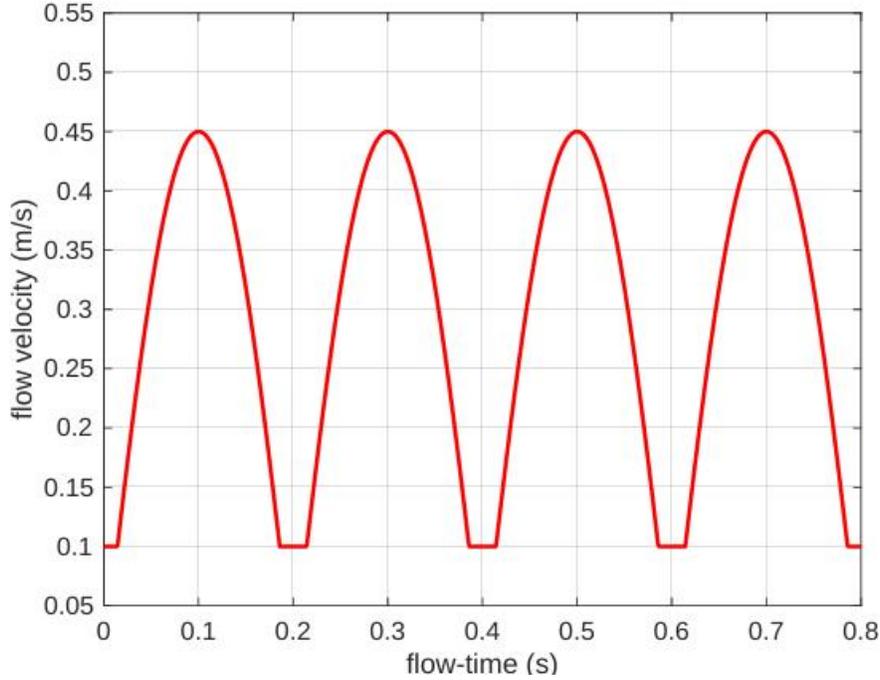

**Figure 4**: Inlet velocity profile for the transient blood flow in all geometries

### 2.5. Numerical solution and simulation

The simulations used the finite-volume discretization method. We applied the semi-implicit (SIMPLE) scheme for velocity-pressure coupling, while the Second Order and Second Order Upwind methods solved the physics of pressure and momentum. We formulated the transient solution using the First Order Implicit Method, setting the time step to 0.004 s with 50 iterations per step [21]. Each simulation ran for approximately 2 minutes on an HP Pavilion 15 X360 PC with an 11th Gen Intel Core™ i7 processor, 16GBRAM, a 1TB SSD, and Windows 11. We obtained and post-processed computational variables, including blood flow velocity, wall pressure distribution, wall shear stress, and their respective volumes as the blood moved through the fluid domain, then compared the results.

### 3. Results and Discussion

In this section, we present the simulation results for blood flow velocity, pressure, and arterial wall shear stress, focusing specifically on the sites of stenosis and the regions upstream to the outlet. The flow velocity is reported using both streamlines and contour plots—this dual representation of hemodynamic properties allows for a more detailed visualization of the flow patterns and potential disturbances caused by the stenosis. Similarly, pressure distribution is presented to show how stenosis-induced narrowing affects pressure gradients, contributing to hemodynamic stress. We further present the arterial wall shear stress, which provide insight into the mechanical forces acting on the arterial walls especially at the site of stenosis.

## 3.1. Flow velocity

### 3.1.1. Velocity streamline

Figure 5 presents the streamline (top) and velocity profiles (bottom, on the YZ plane) for all five geometries, emphasizing the flow behavior at the stenosis region for 10% & 70%, 20% & 60%, 30% & 50%, 40% & 40%, and 0% & 80% configurations, respectively. Blood flow in all geometries follows a similar pattern at the inlet, up to the points of restriction, where distinctive flow patterns emerge. Slight differences appear in the inlet regions, particularly in the intensity of blood flow towards the stenosis throats. In the geometry with 10% & 70% sided stenosis, flow velocity starts relatively low at the entry (0.37 m/s) but stabilizes and increases to 0.56 m/s near the stenosed region. In the 20% & 60% side-stenosed geometry, the flow propagates in four almost equally spaced layers from the entry.

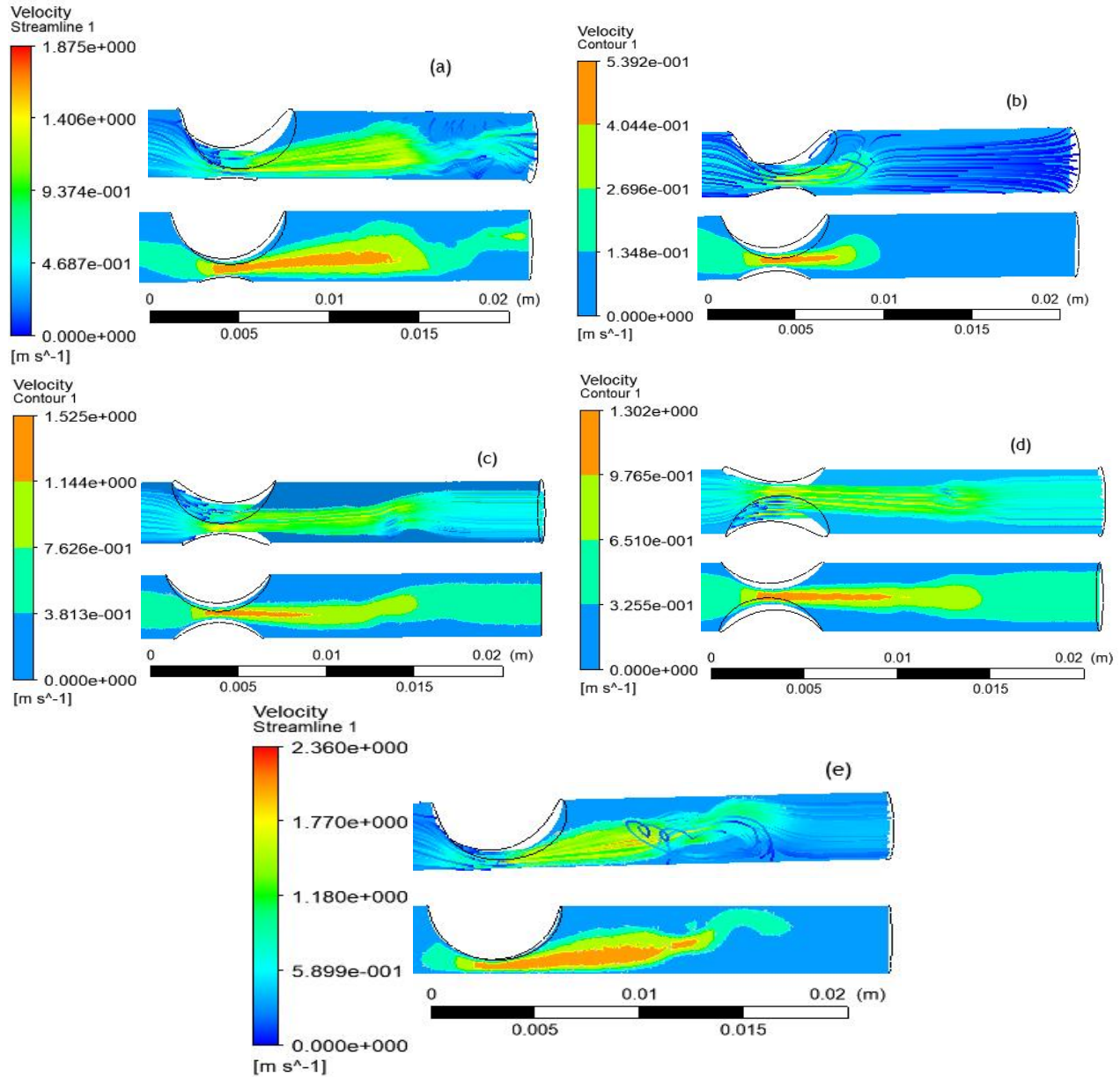

**Figure 5**. Velocity streamline (top) and contour (bottom) at the stenosis constrictions in (a) 10% & 70% (b) 20% & 60% (c) 30% & 50% (d) 40% & 40% and (e) 80% configurations, respectively.

The first layer exhibits the highest magnitude at 0.38 m/s, followed by the second, third, and fourth layers, with magnitudes of 0.32 m/s, 0.27 m/s, and 0.16 m/s, respectively, up to the mid-length of the upstream. The geometry with 30% & 50% sided lesions maintains a uniform flow pattern from the entry to the stenosed region at a velocity of 0.46 m/s, marking the only model with such a consistent flow behavior. In the 40% & 40% sided stenosis configuration, blood enters the inlet with a relatively low velocity range of 0.26–0.39 m/s. However, as the flow develops towards the center, velocity increases to about 0.5 m/s, maintaining steadiness up to the constriction. In contrast, the one-sided 80% stenosis geometry exhibits a completely uniform flow at 0.24 m/s from the inlet. Just before entering the stenosis constriction, the flow velocity rises significantly to 0.71 m/s.

### 3.1.2. Velocity contour comparison

We present a comparative report on the velocity contour (bottom) and the streamlines (top) results shown in Figure 5. In Figure 5a and e, the flow pattern at the constrictions appears similar in the 0% & 80% and 10% & 70%sided stenosed geometries, with flow magnitudes of 2.36 m/s and 1.86 m/s, respectively. After the constrictions, the flow velocity in both geometries decreases continuously, exiting the models at magnitudes of 0.47 m/s and 0.37 m/s, respectively. The 0% & 80% stenosis shows the most significant effect when comparing the two geometries.

In Figure 6b, the 20% & 60% stenosis configuration exhibits an extreme case. This model shows a flow velocity at the constriction throat of less than 1 m/s, the only instance among all models. The velocity drops sharply immediately after the constriction, unlike in other models where it decreases more gradually. The maximum blood flow velocity at the throat reaches 0.54 m/s, and the outlet velocity remains approximately the same as the velocity at the constriction throat. We observe similar blood flow behavior in the 40% & 40% and 30% & 50% sided stenosed geometries (Figure 6c and d). The flow velocity at the constriction throats reaches 1.30 m/s and 1.53 m/s, respectively. The plunging jet-like flow in the 40% & 40% constriction causes a high velocity at the center of the arterial volume, persisting for more than half the length of the outlet region before dropping and exiting the artery. Although the flow behavior is similar in both cases, the delivering velocity magnitude in the 30% & 50% geometry is greater at 0.46 m/s compared to 0.39 m/s in the 40% & 40% stenosed geometry.

As observed in all velocity contours in Figure 6 (bottom), the constriction active radius played a huge role in the velocity profile at the throats in all five geometries. In the one-sided 80% stenosis geometry, where the velocity at the constriction is highest compared to other geometries, the difference in internal arterial wall shape on both ends of the model created the great disruption in flow pattern, influencing the constriction and outlet flow. In general, the one-sided 80% stenosis had the greatest flow restriction effect, followed by the 10% & 70%, 30% & 50%, 40% & 40% and 20% & 60% stenosis configuration, respectively. As seen from the streamlines, the splitting was greatest in the 20% & 60% constriction at the outlet, but at the stenosis, as suggested by the velocity contour, greater vortex was seen in the one-sided 80% stenosis, followed by the 10% & 70% configuration, which began immediately after the constriction and remained relatively unstable up to the outlet. The result shows that flow was more stable in the geometries with the 40% & 40% and 30% & 50% stenosis configuration, implying that in this case, flow dynamics is more unstable in the stenosis configurations with significant difference in ratio of active radius.

### 3.2    Wall pressure

The total pressure magnitudes and distributions at the stenoses for all five geometries are presented in Figure 6. In all geometries, pressure increases upstream of the constriction and drops significantly at the constrictions before rising slightly downstream. The maximum pressure drops at the constriction throats measure -650 Pa, -27 Pa, -440 Pa, -300 Pa, and -880 Pa for the (a) 10% & 70%, (b) 20% & 60%, (c) 30% & 50%, (d) 40% & 40%, and (e) 80% stenosis configurations, respectively. In the 40% & 40% model, the pressure reaches 630 Pa upstream, distal to the constriction, exhibiting distinct behavior compared to the other models. These high-pressure values contribute to the reductions in velocity magnitude at the constriction throats, as shown in Figure 6 (bottom). Notably, wall pressure remains lowest in the 20% & 60% stenosis configuration, which influences the higher velocity observed upstream of the constriction. However, the one-sided 80% stenosis geometry exhibits the greatest pressure drop at the throat, reflecting significant flow restriction along the narrowing base of the artery.

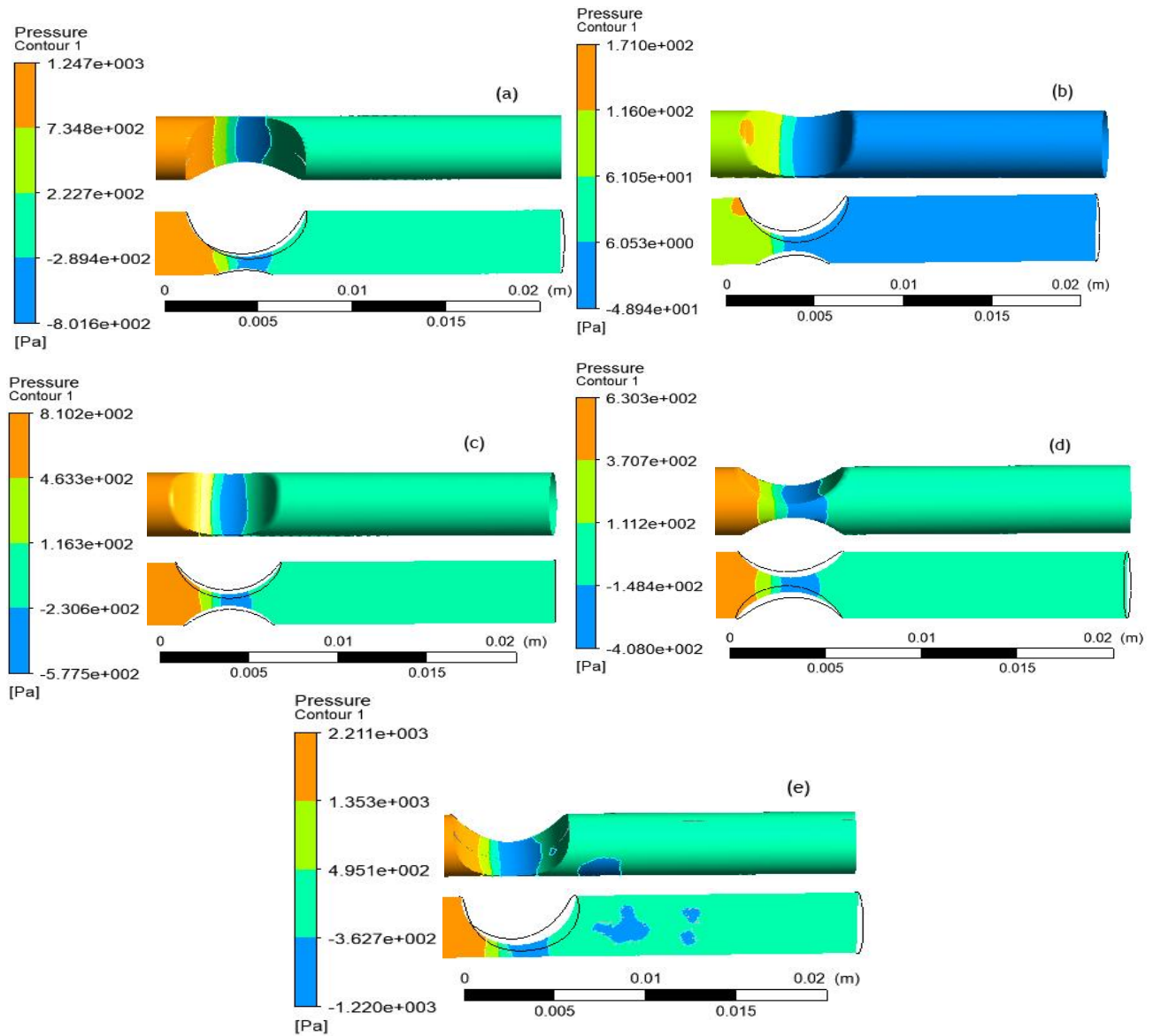

**Figure 6.** Pressure distribution at stenosis site in (a) 10% & 70% (b) 20% & 60% (c) 30% &     50%  (d)  40%  &  40% and (e) 80% configurations, respectively.

### 3.3    Wall shear stress

Figure 7a-e shows the WSS distributions across the geometries. As expected, the maximum WSS occurs at the constrictions in all models, except in the 20% & 60% stenosis configuration. In this case, the stress peaks upstream of the constriction, starting at 1.32 Pa near the entrance and later dropping to 3.36 Pa along the line. Among the models, the one-sided 80% stenosed configuration exhibits the highest WSS at the constriction throat, followed by the 10% & 70%, 30% & 50%, 40% & 40%, and 20% & 60% configurations, respectively. Higher stresses appear downstream in the one-sided 80% and 10% & 70% configurations, caused by turbulent-like flow patterns observed in the streamlines (see Figure 5, top) in these regions. Comparing the stenosis ratios on both sides of each geometry reveals that the smaller stenoses radii (10%, 20%, and 30% in the 10% & 70%, 20% & 60%, and 30% & 50% configurations, respectively) experience more severe WSS effects. Additionally, in these cases, WSS shows a direct relationship with the magnitude of blood flow upstream, at the constriction, and downstream.

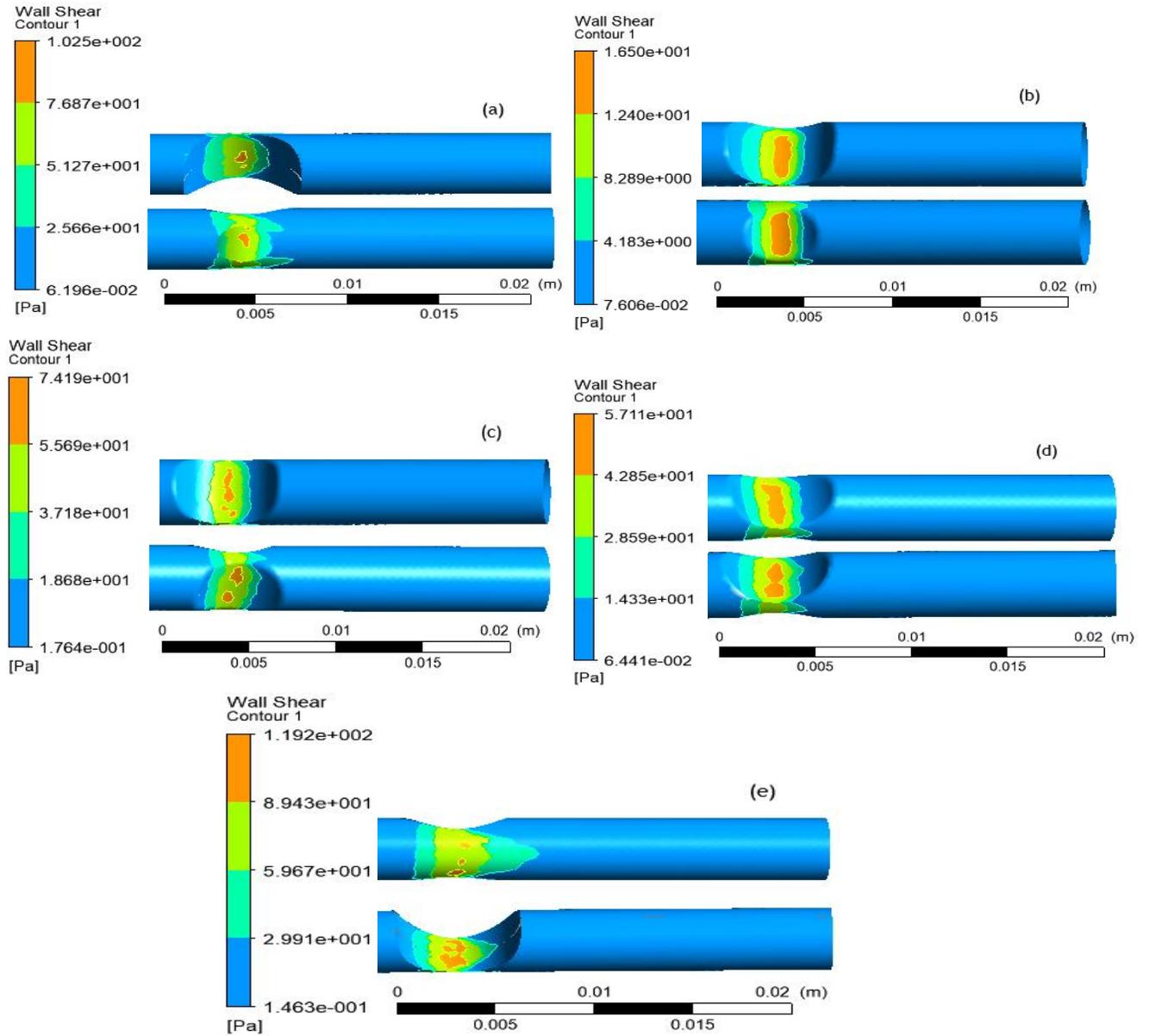

**Figure 7**. Wall shear stress distribution at stenosis sites for  (a) 10% & 70% (b) 20% & 60% (c) 30%  &  50%  (d) 40% & 40% and (e) 80% configurations, respectively.

*3.4      Comparison of overall hemodynamics variables over entire domain*

Figure 8 (A-E) shows the simulation for the full hemodynamic profile for each geometry, highlighting (1) streamlines, (2) velocity contours, (3) wall shear stress, and (4) wall pressure distribution across the entire volume of the stenosed geometries. In Figure 9-11, we presented a comparative hemodynamics chart of the variables across the entire flow domain of each geometries. In Figure 12 and 13, we presented visualization of the maximum hemodynamic variables at the site of stenosis in all geometry utilizing data from the simulation of flow domain, presented in Tables 3 and 4.

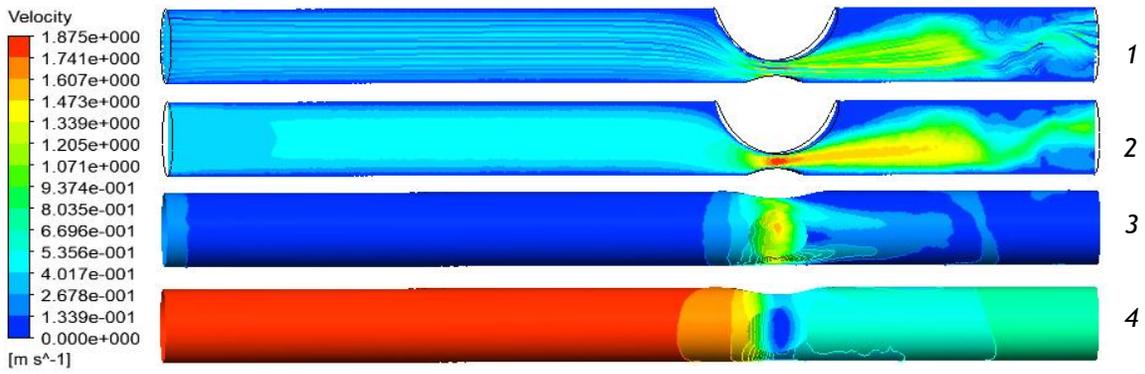

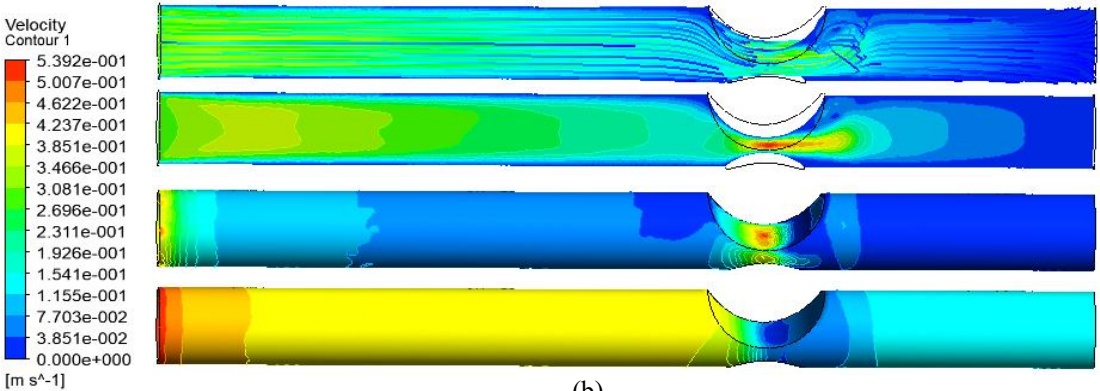

(a)

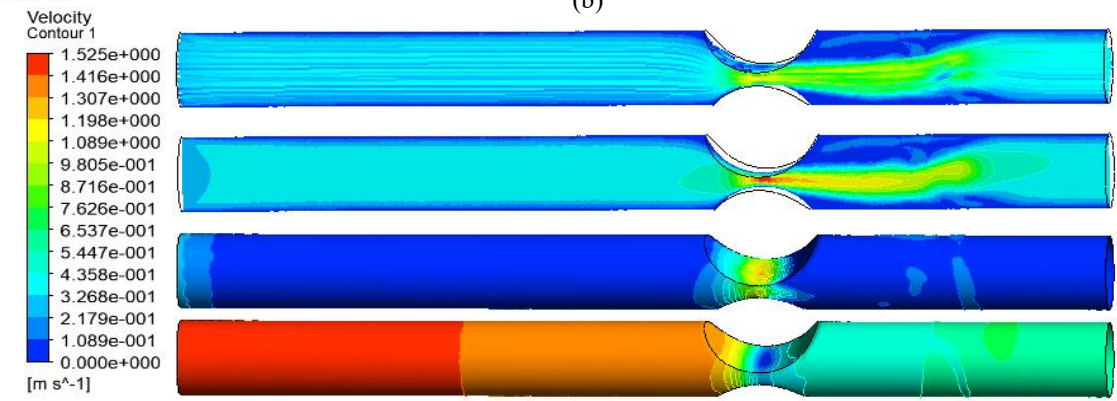

(b)

(c)

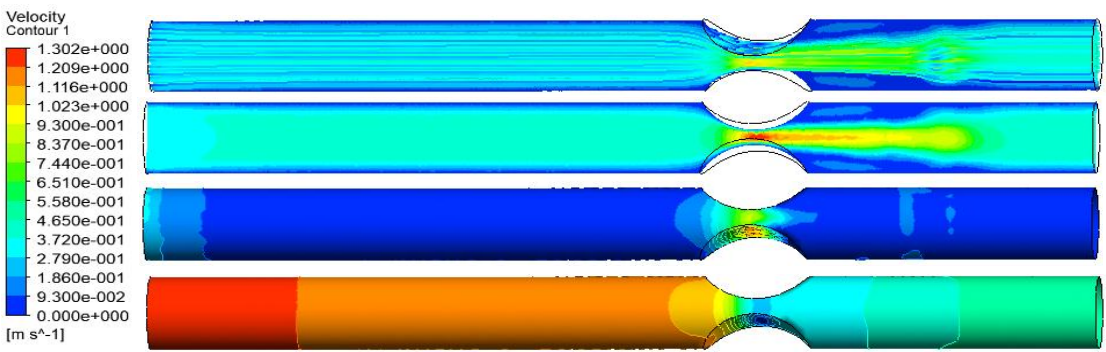

(d)

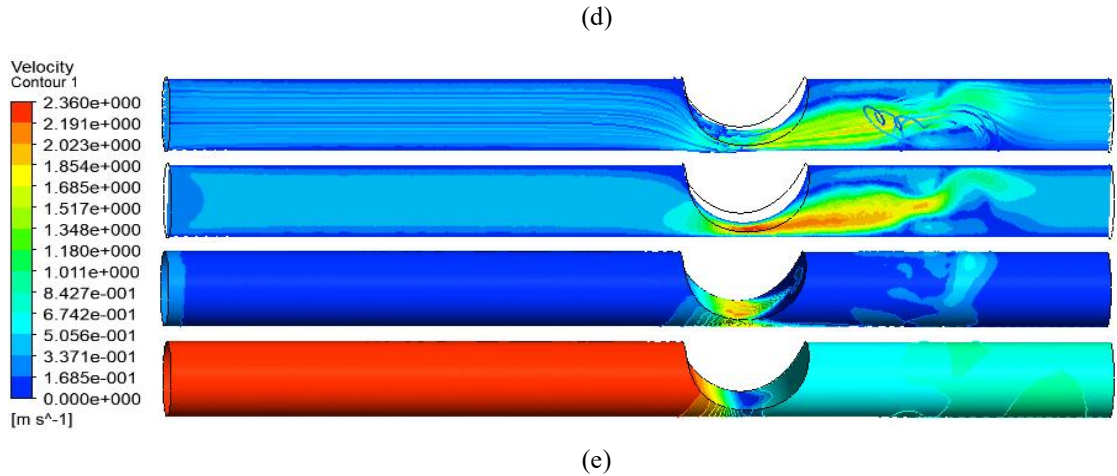

(e)

**Figure 8.** Major computational flow behavior at high values of contour including (1) streamlines, (2) velocity contour, (3) wall shear stress, and (4) wall pressure distribution in geometries with (a) 10% & 70%, (b) 20% & 60%, (c) 30% & 50%, (d) 40% & 40%, and (e) 80% stenosis configurations, respectively.

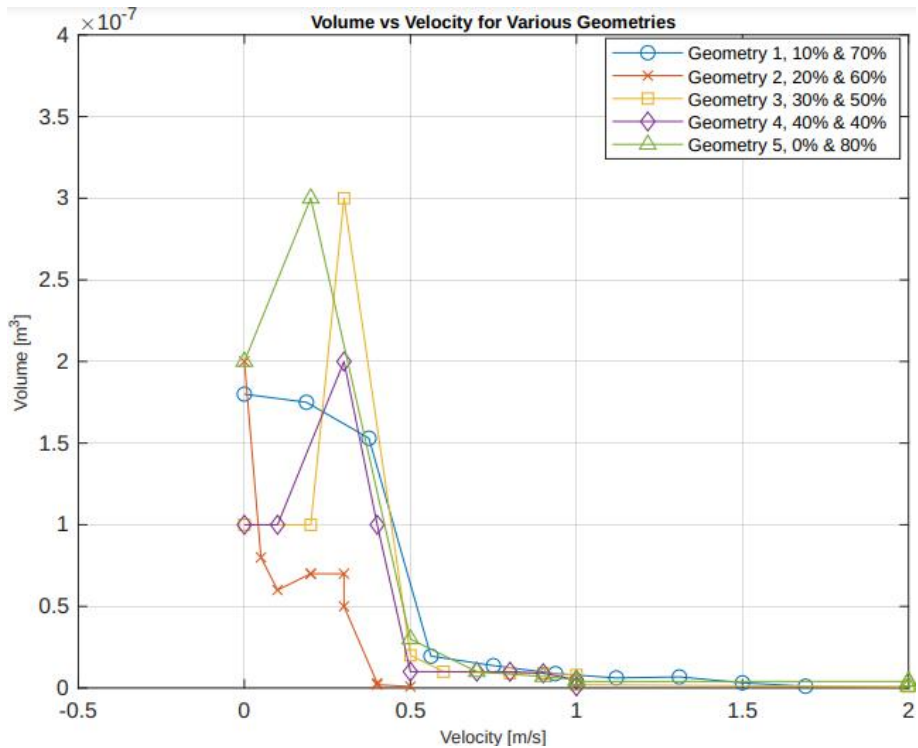

**Figure 9.** Comparative hemodynamic flow velocity through the entire flow domain for each stenosed geometry.

Figure 9 illustrates the relationship between velocity (flow speed) and volume (displacement or flow capacity) across the entire volume of each stenotic geometry. In the 10% & 70% stenosis geometry, velocity increases progressively, peaking at approximately 1.69 m/s, while the corresponding volume drops sharply from $1.80\times10^{-7}$ m³ to nearly $1.09\times10^{-9}$ m³. This trend aligns with the continuity equation in stenotic arteries, where a reduction in cross-sectional area results in increased velocity [25, 26] at the stenosis throat. Post-stenosis, the flow capacity diminishes definitely due to the observed slight turbulence and energy losses. In the 20% & 60% stenosis geometry, velocity reaches a lower peak of around 0.5 m/s, with volumes ranging from $2.00\times10^{-7}$ m3 to $9.00\times10^{-10}$ m³. The volume decreases more gradually compared to the 10% & 70% stenosis case, reflecting less aggressive acceleration due to the moderate stenosis severity. This behavior correlates with findings that asymmetric lesions with moderate constriction still reduce flow but generate fewer disturbances downstream, as reported in previous studies [27], [28]. In the 30% & 50% stenosis configuration, velocity increases significantly, peaking at 2.00 m/s. The volume initially experiences a slight drop but then decreases sharply to $9.00 \times 10{-}10$ m$^3$. The asymmetric distribution of the stenosis amplifies velocity gradients, leading to exacerbated downstream flow disturbances and further reductions in volume. In contrast, the 40% & 40% stenosis configuration exhibits more modest velocities, peaking at 1.00 m/s, with volumes gradually declining from $1.00 \times 10{-}7$ m3 to $1.00 \times 10{-}9$ m$^3$. The symmetric geometry promotes more uniform flow behavior, supporting the idea that symmetric stenoses produce smoother velocity transitions with reduced turbulence compared to asymmetric configurations. In the 0% & 80% stenosis geometry, velocity reaches a peak of 2.00 m/s, accompanied by a sharp drop in volume from $2.00 \times 10{-}7$ m$^3$ to $1.00 \times 10{-}9$ m$^3$. The severe 80% constriction in an asymmetric distribution induces extreme velocity spikes, rapid energy dissipation, and significant flow instability downstream. This behavior aligns with observations of pronounced pressure drops and concentrated wall shear stress (WSS) at the stenosis site, as previously discussed.

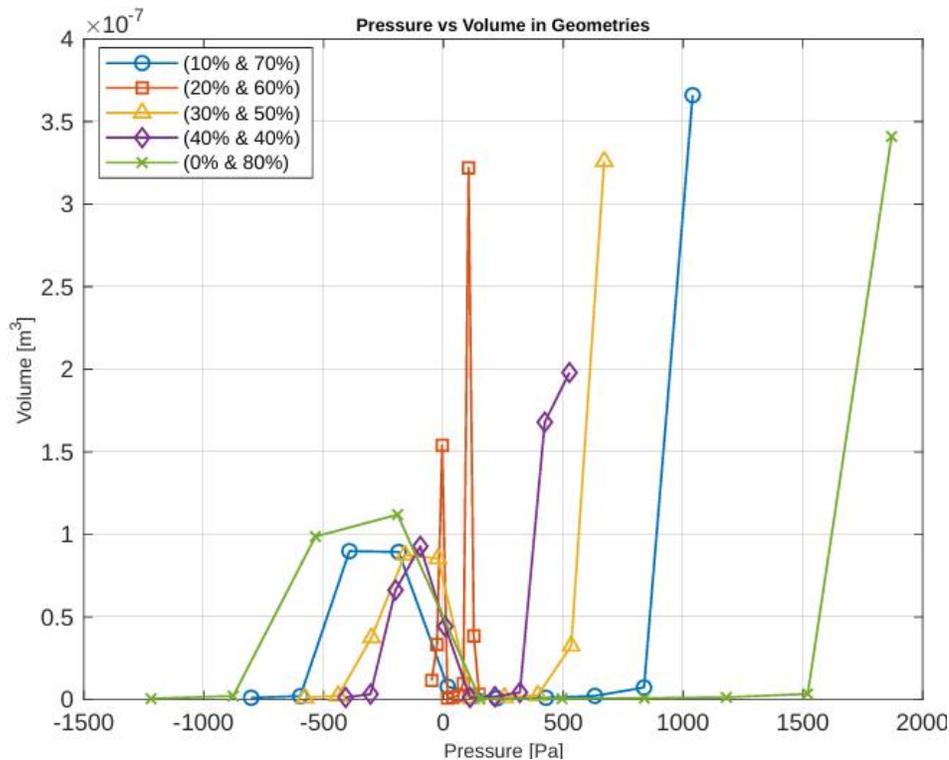

**Figure 10.** Comparative flow pressure against hemodynamic volume through the entire flow domain for each stenosed geometry.

Figure 10 illustrates the variations in volume under different pressures across the five distinct stenotic geometries, each defined by a specific percentage configuration. In Geometry 1 (10% & 70%), the volume remains relatively small under higher negative pressures, indicating strong resistance to compression. As pressure transitions to positive values, the volume exhibits a noticeable non-linear increase, peaking significantly toward the upper end of the range. In Geometry 2 (20% & 60%), the volume shows reduced sensitivity to changes in the negative pressure regime compared to Geometry 1. Positive pressures lead to sharp volume increases, particularly in the mid-range, but the growth stabilizes as pressures reach higher values. This configuration reflects a balance between resistance to compression and expandability under positive pressure. Geometry 3 (30% & 50%) demonstrates a gradual increase in volume across the entire pressure range, suggesting lower elasticity compared to the first two geometries.

The consistent rise in volume for both negative and positive pressures indicates a uniform response, peaking at higher positive pressure values. In Geometry 4 (40% & 40%), the volume responds to pressure changes in a relatively linear manner, without the pronounced peaks observed in other geometries. Negative pressures significantly compress the volume, but the recovery under positive pressure is steady and moderate. This symmetry in geometry likely contributes to the more uniform response. Finally, Geometry 5 (0% & 80%) covers the widest range of pressure values, with sharp volume changes occurring in both the negative and positive regimes. The extreme asymmetry of this geometry results in rapid volume adaptations under varying pressures, reflecting its high sensitivity to arterial deformation.

We observed that all geometries show a non-linear relationship between pressure and volume. The response varies significantly depending on the geometry, with some geometries (e.g., Geometry 1 and 5) being highly sensitive to pressure, while others (e.g., Geometry 3 and 4) show more subdued responses.

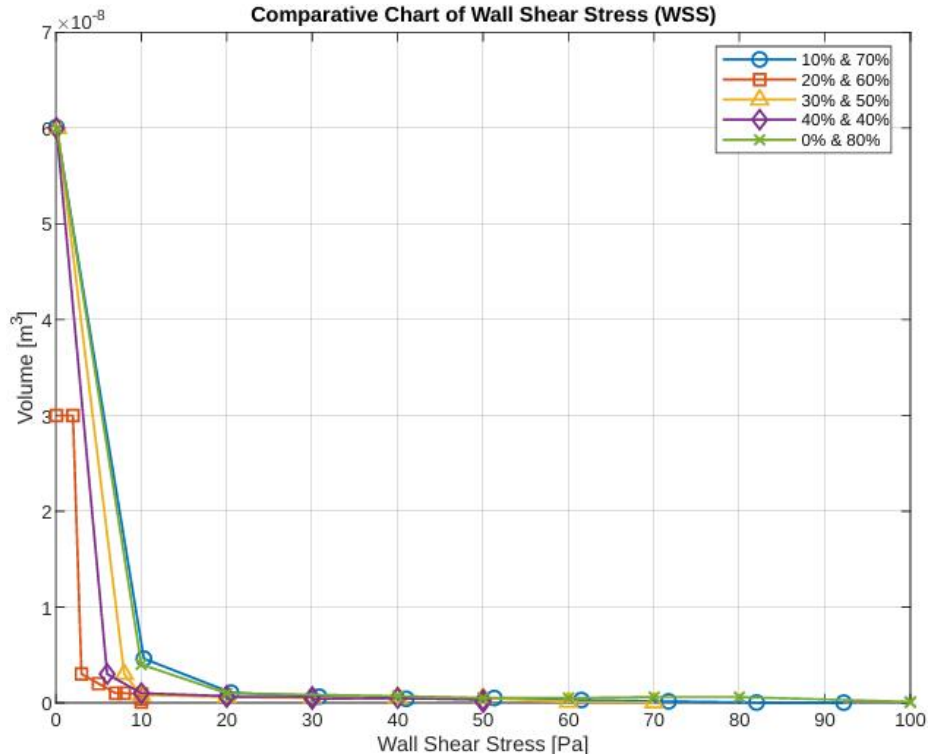

**Figure 11**. Comparative WSS through the entire flow domain for each stenosed geometry.

Figure 11 provides a comparative analysis of WSS across the five geometries, highlighting the relationship between shear forces on the arterial walls and the corresponding volumetric flow distribution. In Geometry 1 (10% & 70%), the volume experiences a sharp decline as WSS increases. At the lowest WSS value of 0.0609 Pa, the volume reaches $6.01 \times 10-8$ m$^3$. However, as WSS increases, the volume decreases drastically, approaching a minimal value of $1.74 \times 10-11$ m$^3$ at 102 Pa. This sharp reduction suggests that higher shear forces create substantial flow resistance, significantly constraining volumetric flow. In Geometry 2 (20% & 60%), WSS starts at 0.08 Pa, with a relatively stable volume of $3.00 \times 10-8$ m$^3$ up to approximately 2 Pa. Beyond this threshold, the volume steadily declines, reaching $8.00 \times 10-11$ m$^3$ at 10 Pa. Compared to Geometry 1, this configuration exhibits a more gradual sensitivity to increasing WSS, indicating a less resistant flow path at lower shear force ranges. The near-constant volume at lower WSS values suggests a reduced flow restriction initially. Geometry 3 (30% & 50%) exhibits a distinct trend. WSS begins at 0.2 Pa, with an initial volume of $6.00 \times 10-8$ m$^3$. As WSS increases, the volume decreases sharply, reaching $2.00 \times 10-11$ m$^3$ at 70 Pa. Interestingly, the volume stabilizes somewhat between 40 Pa and 60 Pa, indicating a potential balance between the imposed shear stress and flow resistance. This behavior suggests that the geometry may permit inconsequential flow distribution within certain stress thresholds.

In Geometry 4 (40% & 40%), WSS starts at 0.06 Pa, and the initial volume matches that of Geometry 3, at $6.00 \times 10-8$ m$^3$. The volume decreases consistently as WSS increase, reaching $7.00 \times 10-11$ m$^3$ at 50 Pa. Like

Geometry 3, the volume stabilizes between 20 Pa and 30 Pa, but the overall decline is less pronounced. This suggests that the symmetric nature of the stenosis in this configuration contributes to smoother flow and reduced vortex. Finally, Geometry 5 (0% & 80%) starts with a WSS of 0.1 Pa and an initial volume of $6.00 \times 10{-}8$ m3. The volume decreases steadily as WSS increases, reaching $4.00 \times 10{-}11$ m3 at 100 Pa. This geometry maintains relatively higher volumes compared to the others at equivalent WSS levels, reflecting its ability to sustain more consistent flow despite the severe stenosis. The high degree of asymmetry likely contributes to this outcome by redistributing shear forces more efficiently.

In a more streamlined evaluation (Table 3 & 4, and Figure 12 & 13), our analysis reveals that stenosis geometry significantly influences the interplay between WSS and volumetric flow. Symmetric geometries like Geometry 4 exhibit smoother transitions and less abrupt declines in volume, while asymmetric configurations like Geometry 1 and Geometry 5 demonstrate heightened sensitivity to shear forces, leading to more pronounced flow constraints.

Table 3. Maximum velocity, pressure, and wall shear stress at the sites of stenosis

|  | 10% & 70% | 20% & 60% | 30% & 50% | 40% & 40% | 0% & 80% |
|---|---|---|---|---|---|
| Max. Vel (m/s) | 1.9 | 0.5 | 1.5 | 1.3 | 2.4 |
| Max. Pre (Pa) | -650 | -27 | -440 | -300 | -880 |
| Max. WSS (Pa) | 89 | 17 | 67 | 51 | 110 |

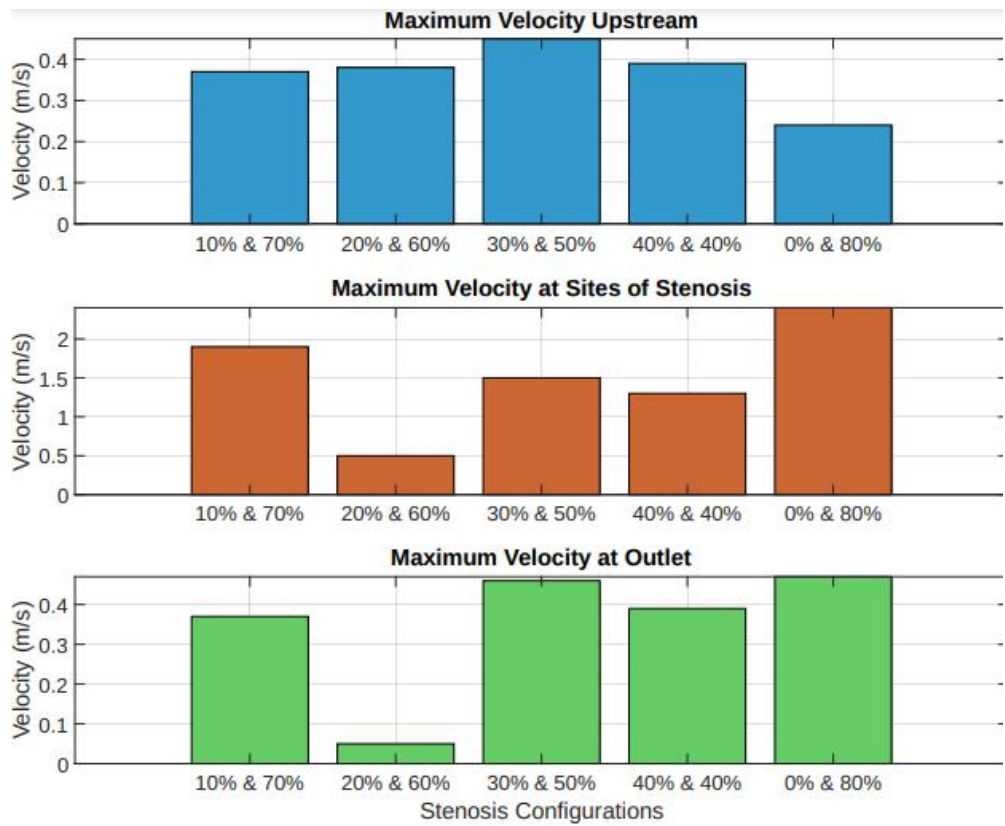

**Figure 12.** Maximum flow velocities in all five geometries at upstream, constriction throats and downstream.

The comparison shows that Geometry 1 exhibits the highest sensitivity to WSS, with a dramatic reduction in volume as WSS increases. In contrast, Geometry 5 shows the least sensitivity, maintaining relatively higher volumes under similar stress conditions. Both Geometry 3 and Geometry 4 exhibit periods of stabilization in volume, suggesting these geometries, though constrained, yield laminar flow through the available passage.

**Table 4.** Maximum velocity of flow upstream, at sites of stenosis and downstream the geometry

|                        | 10% & 70% | 20% & 60% | 30% & 50% | 40% & 40% | 0% & 80% |
|------------------------|-----------|-----------|-----------|-----------|----------|
| Max. Vel. (Upstream)   | 0.37      | 0.38      | 0.45      | 0.39      | 0.24     |
| Max Vel. (Constriction)| 1.9       | 0.5       | 1.5       | 1.3       | 2.4      |
| Max Vel. (Outlet)      | 0.37      | 0.05      | 0.46      | 0.39      | 0.47     |

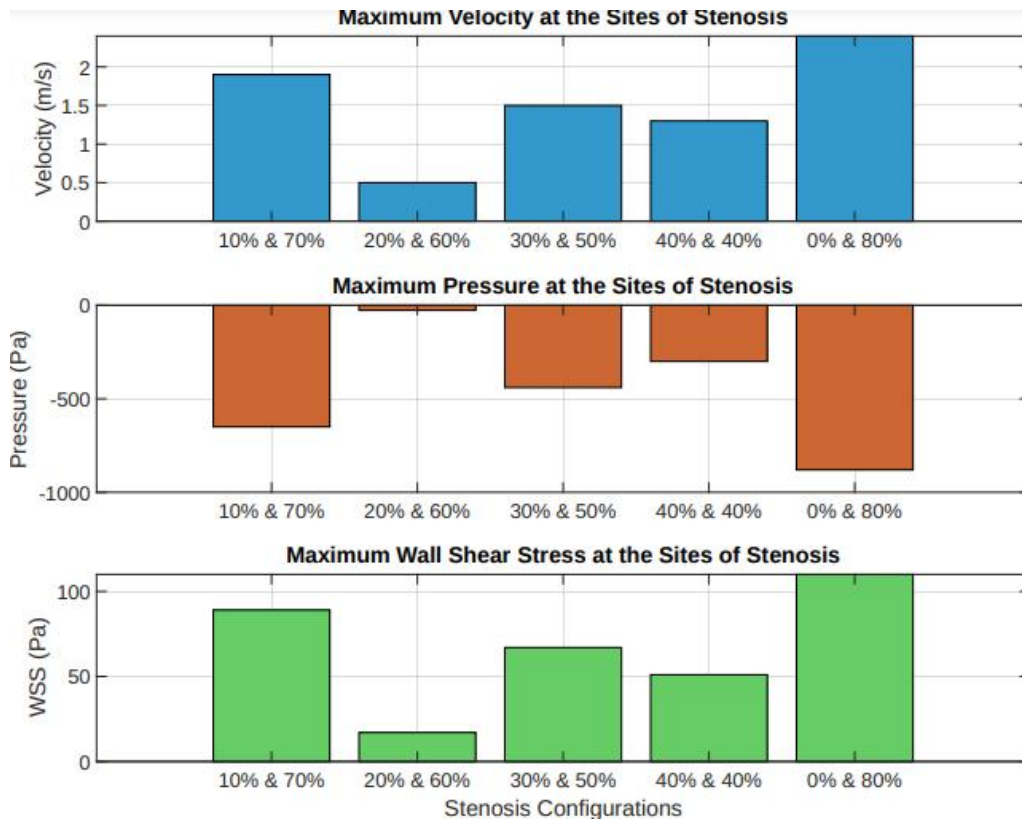

**Figure 13**. Hemodynamic Parameters at the Sites of Stenosis. Direct relationship between the maximum velocities and maximum wall shear stresses at the constriction in all five geometries.

In all of the stenosis configurations, a steady higher velocity was maintained throughout the constriction. It was also observed, how, for example, the 0% & 80% and the 20% & 60% configurations exhibited distinct flow characteristics, including high flow velocities and disrupted flow patterns. This observation implies that cases with such stenosis configurations may be at higher risk of hemodynamic abnormalities and associated complications, perhaps, compared to other forms of stenosis. Furthermore, the observation of higher stresses downstream in the 0% & 80% and 10% & 70% configurations, likely resulting from turbulent-like flow, highlights the importance of considering flow dynamics beyond the constrictions, particularly considering the local flow conditions when assessing the risk of complications, such as thrombus formation or plaque instability that may contribute to regions of low wall shear stress or disturbed flow [4], known to play a role in the development and progression of atherosclerosis.

In addition, the maximum WSS observed at the constrictions indicates areas of high mechanical stress on the arterial walls, which can contribute to endothelial dysfunction, inflammation, and the development of atherosclerotic plaques [29, 30]. These findings imply that clinicians need to be aware of the potential impact of altered flow patterns and increased WSS on downstream segments of the artery, as these factors can affect disease progression and the risk of complications.

Furthermore, the comparison of stenosis ratios on both sides of each geometry reveals that the smaller radii (10%, 20%, and 30% in the 10% & 70%, 20% & 60%, and 30% & 50%, respectively) exhibit more severe effects of WSS. This suggests that the degree of stenosis and the size of the constricted region play a significant role in determining the magnitude of WSS and its impact on arterial hemodynamics, especially in idealistic studies. The results of this study highlight the importance of considering the specific stenosis morphology when evaluating the hemodynamic implications for patients. Variations in flow behavior observed among the different stenosis configurations indicate that different arterial geometries, though of the same asymmetric type lesion (artery reduction), can lead to different flow patterns, velocities, and shear stresses.

## 4. Conclusion

This study investigated how different stenosis configurations affect the hemodynamics of coronary artery models. We aimed to determine whether varying configurations with the same degree of lumen reduction produce similar simulation outcomes, particularly in idealized models commonly used in CFD analyses of coronary studies. We utilized physiological pulsatile flow conditions to analyze five idealized geometries with 10% & 70%, 20% & 60%, 30% & 50%, 40% & 40%, and 0% & 80% stenosis configurations. Our findings highlight that despite the same degree of narrowing, each stenosis configuration notably alters blood flow behavior, pressure distribution, and wall shear stress, resulting in distinct hemodynamic profiles. We observed that asymmetric configurations, particularly those with significant differences in the ratios of active radii, tend to exacerbate flow disruptions and generate higher wall shear stress, which, according to studies, may contribute to endothelial damage and the progression of atherosclerosis. Conversely, symmetric configurations exhibited smoother flow transitions and reduced turbulence, suggesting a potentially less severe impact on arterial health. The results further revealed that one-sided and highly eccentric geometries caused the greatest flow instability, with marked increases in velocity and pressure gradients at the stenosis sites, indicating a higher risk of adverse hemodynamic effects and providing insights into the variability introduced by stenosis morphology.

Our findings challenge the practice of generalizing results across stenosis configurations without accounting for morphological variations, which is prevalent in many CFD studies using idealized models. Previous works often extrapolate findings from single configurations (i.e, 80% stenosis formed by 40 to 40% area reduction), to predict the hemodynamics of different degrees or shapes of such stenosis (80%). However, our study demonstrates that even among configurations with identical lumen reductions, such as 10% & 70% versus 40% & 40%, the resulting hemodynamics differ notably in terms of flow velocity, pressure distribution, and wall shear stress. These differences arise from the unique geometric asymmetries and flow dynamics associated with each configuration. This variability emphasizes the critical importance of avoiding oversimplifications when conducting CFD analyses of stenosed arteries. To ensure accuracy and clinical relevance, future studies should adopt best practices that involve evaluating multiple variations of the same degree of stenosis. Such an approach will capture the full spectrum of possible hemodynamic behaviors and provide a more comprehensive understanding of how arterial morphology influences disease progression and treatment outcomes. One potential benefit of our findings and subsequent studies is improved patient-specific diagnostics, and treatment planning.

In future work, we aim to extend this analysis to patient-specific coronary geometries with this lesion variation, and incorporate fluid-structure interactions to simulate the dynamic behavior of arterial walls. Through this study, we will further establish the efficacy of none dynamic arterial walls in studies of coronary artery stenosis. Additionally, we believe that the applicability of this research could be enhanced through the integration of machine learning techniques. Future research can use this techniques to validate cases in the current study.

## Authors Contribution

AJO: Conceptualization, design, methodology, CAD models, simulations and result post-processing. AJO AAA and ODA: Result analysis, data presentation and writing of initial manuscript. EOI: Supervision, editing and reviewing of finial manuscript. All authors read and approved the final manuscript.

## Ethical approval
Not required

## Funding
None

## Competing Interest
None

## References

1] Bego˜na Lavin Plaza, Iakovos Theodoulou, Imran Rashid, Reza Hajhosseiny, Alkystis Phinikaridou, and Rene M Botnar. Molecular imaging in ischemic heart disease. Current Cardiovascular Imaging Reports, 12:1–12, 2019

[2] Kouichi Ozaki and Toshihiro Tanaka. Molecular genetics of coronary artery disease. Journal of human genetics, 61(1):71–77, 2016.

[3] Eduardo Pozo, Pilar Agudo-Quilez, Antonio Rojas-Gonz´alez, Teresa Alvarado, Mar´ıa Jos´e Olivera, Luis Jes´us Jim´enez-Borreguero, and Fernando Alfonso. Noninvasive diagnosis of vulnerable coronary plaque. World Journal of Cardiology, 8(9):520, 2016

[4] Ayodele James Oyejide, Adetokunbo Andrew Awonusi, and Ebenezer Olubunmi Ige. Fluid-structure interaction study of hemodynamics and its biomechanical influence on carotid artery atherosclerotic plaque deposits. Medical Engineering & Physics, 117:103998, 2023.

[5] Samson Kolawole Fasogbon, Funmilayo Helen Oyelami, Emmanuel Oreoluwa Adetimirin, and Ebenezer Olubunmi Ige. On blasius plate solution of particle dispersion and deposition in human respiratory track. Mathematical Modelling of Engineering Problems, 6(3), 2019.

[6] Lei Fan, Haifeng Wang, Ghassan S Kassab, and Lik Chuan Lee. Review of cardiac–coronary interaction and insights from mathematical modeling. WIRES Mechanisms of Disease, 16(3):e1642, 2024.

[7] Violeta Carvalho, Diana Pinho, Rui A Lima, Jos´e Carlos Teixeira, and Senhorinha Teixeira. Blood flow modeling in coronary arteries: A review. Fluids, 6(2):53, 2021.

[8] AJ Oyejide, E Emmanuel, AA Awonusi, and EO Ige. A computational study of respiratory biomechanics in idealized healthy and stenosed subsegmental bronchi section of infant, child and adult airways. Series on Biomechanics, 2022.

[9] Oyejide James Ayodele, Atoyebi Ebenezer Oluwatosin, Olutosoye Christian Taiwo, and Ademola Adebukola Dare. Computational fluid dynamics modeling in respiratory airways obstruction: current applications and prospects. Int J Biomed Sci Eng., 9(02):16, 2021.

[10] Ayodele James Oyejide, Chidera Samuella Okeke, Jesuloluwa Emmanuel Zaccheus, and Ebenezer Olubunmi Ige. Computational assessment of turbulent eddy impact on hydrodynamic mixing in a stirred tank bioreactor with vent based impellers. arXiv preprint arXiv:2412.18660, 2024.

[11] Boyang Su, Jun-Mei Zhang, Hua Zou, Dhanjoo Ghista, Thu Thao Le, and Calvin Chin. Generating wall shear stress for coronary artery in real-time using neural networks: Feasibility and initial results based on idealized models. Computers in biology and medicine, 126:104038, 2020.

[12] Masoud Ahmadi and Reza Ansari. Computational simulation of an artery narrowed by plaque using 3d fsi method: influence of the plaque angle, non-newtonian properties of the blood flow and the hyperelastic artery models. Biomedical Physics & Engineering Express, 5(4):045037, 2019.

[13] Violeta Carvalho, Diogo Lopes, Joˆao Silva, Hˊelder Puga, Rui A Lima, Josˊe Carlos Teixeira, and Senhorinha Teixeira. Comparison of cfd and fsi simulations of blood flow in stenotic coronary arteries. In Applications of Computational Fluid Dynamics Simulation and Modeling. IntechOpen, 2022.

[14] Thanapong Chaichana, Zhonghua Sun, and James Jewkes. Computational fluid dynamics analysis of the effect of plaques in the left coronary artery. Computational and mathematical methods in medicine, 2012(1):504367, 2012.

[15] Sarfaraz Kamangar, Govindaraju Kalimuthu, Irfan Anjum Badruddin, A Badarudin, NJ Salman Ahmed, and TM Yunus Khan. Numerical investigation of the effect of stenosis geometry on the coronary diagnostic parameters. The Scientific World Journal, 2014(1):354946, 2014.

[16] Prachi D Dwidmuthe, Gaurav G Dastane, Channamallikarjun S Mathpati, and Jyeshtharaj B Joshi. Study of blood flow in stenosed artery model using computational fluid dynamics and response surface methodology. The Canadian Journal of Chemical Engineering, 99:S820–S837, 2021.

[17] Navid Freidoonimehr, Rey Chin, Anthony Zander, and Maziar Arjomandi. Effect of shape of the stenosis on the hemodynamics of a stenosed coronary artery. Physics of Fluids, 33(8), 2021.

[18] Md Alamgir Kabir, Md Ferdous Alam, and Md Ashraf Uddin. Numerical simulation of pulsatile blood flow: a study with normal artery, and arteries with single and multiple stenosis. Journal of Engineering and Applied Science, 68:1–15, 2021.

[19] Mehdi Jahangiri, Mohsen Saghafian, and Mahmood Reza Sadeghi. Numerical simulation of hemodynamic parameters of turbulent and pulsatile blood flow in flexible artery with single and double stenoses. Journal of Mechanical Science and Technology, 29:3549–3560, 2015.

[20] SM Abdul Khader, Raghuvir Pai, Ganesh Kamath, Mohammed Zuber, and VRK Rao. Haemodynamic study of flow in concentric and eccentric stenosed carotid artery. In MATEC Web of Conferences, volume 144, page 01024. EDP Sciences, 2018.

[21] Micha l Tomaszewski, Pawe l Baranowski, Jerzy Ma lachowski, Krzysztof Damaziak, and Jakub Buka la. Analysis of artery blood flow before and after angioplasty. In AIP Conference Proceedings, volume 1922. AIP Publishing, 2018.

[22] Md Foysal Rabbi, Fahmida S Laboni, and M Tarik Arafat. Computational analysis of the coronary artery hemodynamics with different anatomical variations. Informatics in Medicine Unlocked, 19:100314, 2020.

[23] Mariia Timofeeva, Andrew Ooi, Eric KW Poon, and Peter Barlis. Numerical simulation of the blood flow through the coronary artery stenosis: effects of varying eccentricity. Computers in Biology and Medicine, 146:105672, 2022.

[24] Ayodele J Oyejide et al. Simulation of inspiratory airflow in stenotic trachea and its effect on mainstem bifurcation. Saudi J Biomed Res, 6(11):256–263, 2021.

[25] Carlo Di Mario, Nicolas Meneveau, Robert Gil, Peter de Jaegere, Pim J de Feyter, Cornelis J Slager, Jos RTC Roelandt, and Patrick W Serruys. Maximal blood flow velocity in severe coronary stenoses measured with a doppler guidewire: limitations for the application of the continuity equation in the assessment of stenosis severity. The American journal of cardiology, 71(14):D54–D61, 1993.

[26] Hamidreza Savalanpour, Bijan Farhanieh, and Hossein Afshin. Proposing a general formula to calculate the critical velocities in tunnels with different cross-sectional shapes. Tunnelling and Underground Space Technology, 110:103798, 2021.

[27] Gilwoo Choi, Joo Myung Lee, Hyun-Jin Kim, Jun-Bean Park, Sethuraman Sankaran, Hiromasa Otake, Joon-Hyung Doh, Chang-Wook Nam, Eun-Seok Shin, Charles A Taylor, et al. Coronary artery axial plaque stress and its relationship with lesion geometry: application of computational fluid dynamics to coronary ct angiography. Cardiovascular Imaging, 8(10):1156–1166, 2015.

[28] Kalimuthu Govindaraju, Irfan Anjum Badruddin, Girish N Viswanathan, SV Ramesh, and A Badarudin. Evaluation of functional severity of coronary artery disease and fluid dynamics' influence on hemodynamic parameters: A review. Physica Medica, 29(3):225–232, 2013.

[29] Manli Zhou, Yunfeng Yu, Ruiyi Chen, Xingci Liu, Yilei Hu, Zhiyan Ma, Lingwei Gao, Weixiong Jian, and Liping Wang. Wall shear stress and its role in atherosclerosis. Frontiers in cardiovascular medicine, 10:1083547, 2023.

[30] Habib Samady, Parham Eshtehardi, Michael C McDaniel, Jin Suo, Saurabh S Dhawan, Charles Maynard, Lucas H Timmins, Arshed A Quyyumi, and Don P Giddens. Coronary artery wall shear stress is associated with progression and transformation of atherosclerotic plaque and arterial remodeling in patients with coronary artery disease. Circulation, 124(7):779–788, 2011.